# Quantum cascade phenomenon in natural atomic superlattices formed in $Bi_2Sr_2CaCu_2O_8$ single crystals

Vladimir M. Krasnov

*Department of Physics, Stockholm University, AlbaNova University Center, SE-10691*

*Stockholm, Sweden*

**Abstract:**

Natural atomic superlattices are formed in certain strongly anisotropic layered compound. Here we study interlayer transport in single crystals of $Bi_2Sr_2CaCu_2O_{8+\delta}$ cuprates, which represent stacks of atomic scale "intrinsic" Josephson junctions. A series of resonant dips in conductance is observed at condition when bremsstrahlung and recombination bands in non-equilibrium spectrum of Josephson junctions overlap. The phenomenon is explained in terms of self-detection of a new type of collective strongly non-equilibrium state in stacked Josephson junctions, bearing certain resemblance with operation of a Quantum Cascade Laser. Conclusions are supported by in-situ generation-detection experiments, providing evidence for cascade amplification of recombination radiation, and by numerical simulations.

Superlattices have become an essential part of modern electronics and optics (1,2). One of superlattice-based quantum devices is a Quantum Cascade Laser (QCL) (1), in which superlattice forms a stack of tunnel junctions. Sequential tunneling in the stack both creates a non-equilibrium electron distribution (population inversion) and facilitates *cascade amplification* of non-equilibrium photon distribution. Both processes are central to laser action. While fabrication of homogeneous superlattices is a formidable task (3), atomically perfect superlattices are naturally formed in certain layered compounds (4,5,6,7). Properties of those compounds can be varied by intercalation (6,7,8) and doping (9), allowing flexibility, similar to band engineering in semiconductor superlattices (1,3). Fig. 1 a) shows the crystallographic structure of $Bi_2Sr_2CaCu_2O_{8+\delta}$ (Bi-2212) high temperature superconductor (HTSC), which represents a natural stack of atomic scale "intrinsic" Josephson junctions (IJJ's) (10,11,12,13,14).

Here we study interlayer transport in small mesa structures micro/nano-fabricated on top of Bi-2212 single crystals. A new phenomenon, seen as a series of resonant dips in interlayer conductance at multiples of the sum-gap voltage is observed. It is concluded that the phenomenon represents a new type of collective strongly non-equilibrium state in stacked IJJ's, bearing certain resemblance with operation of the QCL. Conclusions are supported by in-situ generation-detection experiments, providing evidence for cascade amplification of recombination radiation in stacked IJJ's, and by numerical simulations.

To study interlayer transport, small mesa structures containing few atomic layers were micro/nano-fabricated on top of Bi-2212 single crystals. Fig 1 b) represents sketch of a triple-mesa structure, consisting of a generator (B) and a detector (C) mesas on top of a common pedestal (D), and several single mesas (A, E) required for independent biasing of the generator and the detector. Mesas B and C were made by cutting the initial mesa by Focused Ion Beam. This allows nano-scale separation between mesas B and C, needed for in-situ detection (15). Details of sample fabrication are described elsewhere (14).

Fig. 1 c) shows a Current-Voltage characteristic (IVC) at $T$=4.2K for a single mesa on a slightly overdoped crystal with $T_c$ = 92.5 K. Multi-branch structure is due to one-by-one switching of IJJ's from the superconducting into the resistive state. Figs. 1 d and e) shows IVC's of a triple-mesa fabricated on a near optimally doped crystal, $T_c$=93.6 K measured in three probe ($I^+= V^+$= B, $I^-$ = A, $V^-$ = E) and four probe ($I^+$=B, $V^+$=C, $I^-$=A, $V^-$=E, see Fig 1 b) configurations, respectively. The pedestal mesa D had an area ~4x5 μm$^2$, mesas B and C had approximately equal areas ~ 4x2.25 μm$^2$, and were separated by ~0.5 μm. In the three probe configuration the IVC is a sum of IVC's of mesas B and D, as seen from appearance of two families of branches with a difference in critical currents determined by the ratio of areas of mesas B and D ~0.45.

Fig. 2 a) shows tunneling conductance $\sigma = dI/dV(V)$ at $T<T_c$. A smooth normal state characteristic at $T$ = 100 K was subtracted to improve the spectroscopic resolution (9). IVC's of single mesas have the shape typical for Superconductor-Insulator-Superconductor (SIS) tunnel junctions (8,9,13) with a sharp peak in conductance at the sum-gap voltage $V_g$ =2$\Delta$/$e$ per junction, where $\Delta$ is the superconducting energy gap. Improved resolution allows for the first time observation of a fine structure at high bias: arrows in Fig. 2 a) indicate appearance of dips in d$I$/d$V$ at $V>V_g$. Those features can be clearly seen in the second derivative plot d$^2I$/d$V^2$($V$), shown in Fig. 2 b) for a slightly underdoped mesa, $T_c$ = 88 K. Here up to three kinks above $V_g$ can be distinguished. Figs. 2 c,d) show $T$-dependencies of the peak and the dip voltages per IJJ. In Fig. 2d) data for another mesa on the same crystal are also shown to demonstrate reproducibility of the features. Lines in Fig. 2 c) indicate that the two consecutive dips in d$I$/d$V$ occur at $V$=2$V_g$=4$\Delta$/$e$ and 3$V_g$ =6$\Delta$/$e$ per junction, respectively. In Fig. 2 d) up to three dips are observed in a wide $T$-range. Here dips occur at $V$ slightly smaller than multiples of $V_g$: by 10% for the line ~4$\Delta$/$e$, and by 15% for the lines ~6$\Delta$/$e$ and ~8$\Delta$/$e$. Such deviation can be attributed to self-heating at large bias (15).

So far peculiarities at multiples of $V_g$ were not reported neither for single nor stacked Josephson junctions. On the other hand, peculiarities in non-equilibrium phonon distribution at the corresponding voltages have been observed in phonon generation-detection experiments (16,17,18,19,20,21). In that case a second SIS junction (the detector) was connected to the biased junction (the generator). The quasiparticle (QP) conductance of the detector increased stepwise at $V = nV_g$ in the generator, due to excess flow of non-equilibrium phonons with energy $\Omega \geq 2\Delta$, sufficient for Cooper pair breaking in the detector. However, no peculiarities in the generator itself were reported. Also conductance in Fig. 2 a) exhibits resonant dips (rather than steps), which at least partially recover at larger voltages. This difference is related to the specific sample geometry: in previous experiments the generator was biased independently from the detector and always remained in a quasi-equilibrium state. To the contrary, in our case the signal is due to collective self-detection in simultaneously biased stacked junctions, which all may be in a strongly non-equilibrium state.

To understand non-equilibrium processes in stacked Josephson junctions, let's consider the energy diagram in Fig. 3 a). Tunneling leads to non-equilibrium population of QP's with the maximum at $E = eV$-$\Delta$. The decay of QP's typically follows a two-step process (20) with the QP first relaxing to the bottom of the empty band, emitting a bremsstrahlung (braking) radiation (17). At the second stage, two QP's from the bottom of the band recombine into the Cooper pair, emitting recombination radiation. The process is repeated in the next junction leading to cascade amplification of radiation in the stack, as illustrated by wavy arrows in Fig. 3 a).

Emission of a boson with frequency $\Omega$ due to relaxation of QP with energy $E$ occurs at the rate

$$\Gamma(\Omega, E) \propto D(\Omega)\rho(E)\rho(E-\Omega)f(E)[1-f(E-\Omega)][1+g(\Omega)], \qquad (1)$$

where $D$ and $\rho$ are densities of states, and $g$ and $f$ are distribution functions for bosons and QP's, correspondingly. Factors 1 and $g(\Omega)$ in the last term of Eq. 1 represent probabilities of spontaneous and stimulated emissions, respectively.

Fig. 3 b) shows calculated spectrum of non-equilibrium phonons for a single junctions at $V=3.5\Delta/e$ (see the Appendix for details). Two discrete phonon bands with abrupt cutoffs at $\Omega_B$ $eV-2\Delta$ for braking and $\Omega_R \geq 2\Delta$ for recombination bands are seen. Spectra are peaked at the band edges, indicating the two most probable QP decaying events, described above. At $V \geq 4\Delta/e$ braking and recombination bands overlap. Now high frequency braking phonons with $\Omega_B \geq 2\Delta$ can split Cooper pairs and generate secondary QP's. Relaxation of secondary QP's leads to appearance of secondary braking bands with

$$\Omega_B \quad eV - 2n\Delta, n=1,2,3,\ldots, \qquad (2)$$

where $n$-1 is a number of secondary relaxation stages (20).

From Figs. 2 c,d) and Eq. 2 it follows that the dips occur upon collapse of the recombination band with one of the braking bands. According to Eq. 1, collapse of the two bands accelerates emission at the overlap frequency, $\Omega = 2\Delta$, due to stimulated emission. But stimulated emission is significant only if strongly non-equilibrium population $g(\Omega)$ was initially present in the bands. Normally, this does not occur in single Josephson junctions. However, in stacked junctions strong disequilibrium can be achieved due to cascade amplification of non-equilibrium radiation, as shown by wavy arrows in Fig. 3 a). The cascade amplification is efficient only if junctions in the stack are identical. Indeed, the dips were pronounced only in mesas with identical junctions. High uniformity of our mesas can be seen from the periodicity of multi-branch IVC in Fig. 1. It can also be seen from the sharpness of the superconducting peak in Fig. 2 a), which indicates that all junctions reach $V_g$ at the same current. In experiment, a correlation between the height of the peak at $V_g$ and the depth of the dip at $2V_g$ was observed.

To independently study non-equilibrium radiation from the stack, generation-detection experiments were carried out using triple-mesa structures, shown in Fig 1 b). Fig. 4 a) shows a four-probe IVC at $T=75.2$K for the same triple-mesa D as in Fig. 1 e) and two IVC's at close $T$ for a single mesa A (~4x2.25 $\mu m^2$) on the same crystal. It is seen that the IVC's of the single mesa exhibit the conventional kink at the sum-gap bias. In contrast, the IVC of the triple-mesa suddenly drops to a smaller voltage at $I$~0.7mA. The curve in the middle of Fig. 1 a) shows the same IVC, in which $I$ and $V$ were scaled by area and the number of IJJ's so that it can be compared with the IVC's of mesa A. It is seen that at low bias the scaled IVC coincides with the IVC of mesa A at $T=$ 74.9K. At larger bias (after the drop) the shape of the IVC is similar to the IVC of the mesa A at $T=$79.3K, indicating a minor self-heating from the mesa B (15). Both observations confirm the validity of the scaling and show that the genuine IVC's are determined solely by areas of IJJ's. From comparison of the IVC's it becomes obvious that the voltage drop occurs when IJJ's in mesa D reach the sum-gap bias, as shown by the arrow in Fig. 4 a).

Fig. 4 b) shows four-probe $dI/dV(I)$ characteristics for the same triple-mesa at $T$=65K. Here correlation between the voltage drop ($dI/dV \approx 0$ at 1.1mA<$I$<1.2 mA) and the peak at the sum-gap bias, $I_p$, in mesa D can be seen explicitly. Noticeably, an additional peak, never observed in single mesa characteristics, occurs at $I=I_p$*0.45, corresponding to the sum-gap bias in the mesa B.

Thus, a specific response in the detector mesa C occurs when IJJ's in both generator mesas B and D reach the sum-gap bias. It should be emphasized that the response is not due to QP injection because the mesa C is unbiased and the QP current in mesas B and D changes gradually through the singularities. Therefore, the observed singularities indicate appearance of strong recombination radiation in IJJ's of generator mesas B and D when they reach $V_g$. This recombination radiation, $\Omega_R \geq 2\Delta$, results in partial depairing and increase of QP population in mesa D, which is sensed by the detector. Since only the generator D is in direct contact with the detector C the response from D is stronger (discontinuity in the IVC) than from B (peak in the derivative). The response can be attributed to appearance of in-plane resistance between mesas B and C (22), and to disbalance in QP population between mesas D and C (23), leading to appearance of negative voltage at mesa C similar in origin to the built-in potential in $n$-$p$ junctions. Both types of responses enhance with increasing depairing in mesa D.

A closer examination of triple mesa IVC's reveals that the voltage drop consists of a set of tiny sub-branches, see Fig. 4 c). Their number is approximately equal to the number of IJJ's in the mesa D. This indicates that the non-equilibrium radiation in the stack D is amplified in a cascade manner each time an additional IJJ in mesa D reaches $V_g$.

To summarize, we observed a new phenomenon in single crystals of layered Bi-2212 cuprates. It is seen as a set of resonant dips in c-axis conductance at multiples of the sum-gap voltage $2(n+1)\Delta/e$, $n$=1,2,3, at which bremsstrahlung and recombination bands in the spectrum of Josephson junctions overlap. In-situ generation-detection experiments provided evidence for strong recombination radiation and for cascade amplification of radiation in the stack of atomic scale intrinsic Josephsons junctions, naturally formed in Bi-2212 crystals. This brings us to conclusion that the phenomenon represents a new type of collective nonequilibrium state in the natural atomic superlattice, which bears certain resemblance with operation of the QCL. As in the QCL, the nonequilibrium QP distribution is created by interlayer tunneling and stimulated emission is achieved by cascade amplification of radiation upon sequential tunneling in the stack. On the other hand, emission in Bi-2212 is conceptually different from the QCL: (i) Emitted is not light but other bosons (e.g., phonons or spin waves), coupled to superconducting QP's. (ii) The band gap is not due to level quantization in a quantum well, but is the superconducting energy gap. Whence, the effect disappears at $T > T_c$. (iii) Stimulated emission occurs when two bosonic, rather than electronic, bands overlap. Since the overlap frequency, $\Omega$=2$\Delta$, is sufficient for pair braking, it leads to catastrophic, resonant suppression of superconductivity, resulting in the drop of conductance to approximately the normal state value, as seen from Fig. 2 a).

Our observation has several interesting implications:

(i) It can be important for understanding HTSC. Bosonic excitations mediating superconductivity are emitted upon recombination of nonequilibrium QP's (eg., phonons are emitted in conventional superconductors). Therefore, identification of non-equilibrium radiation emitted by IJJ's may disclose the coupling mechanism of HTSC.

(ii) So far superlattice devices relied on complicated fabrication of artificial multilayers. Our data shows that similar and even more unusual physical phenomena may occur in atomic superlattices naturally formed in a variety of layered compounds. This may open a possibility for building novel electronic devices at the ultimate atomic scale.

**Acknowledgments**

I am grateful to I.Zogaj for assistance with fabrication of triple-mesa structures and to A.Golubov and A.Kozorezov for fruitful discussions.

The work was supported by Knut and Alice Wallenberg foundation and Swedish research council.

**Figures**

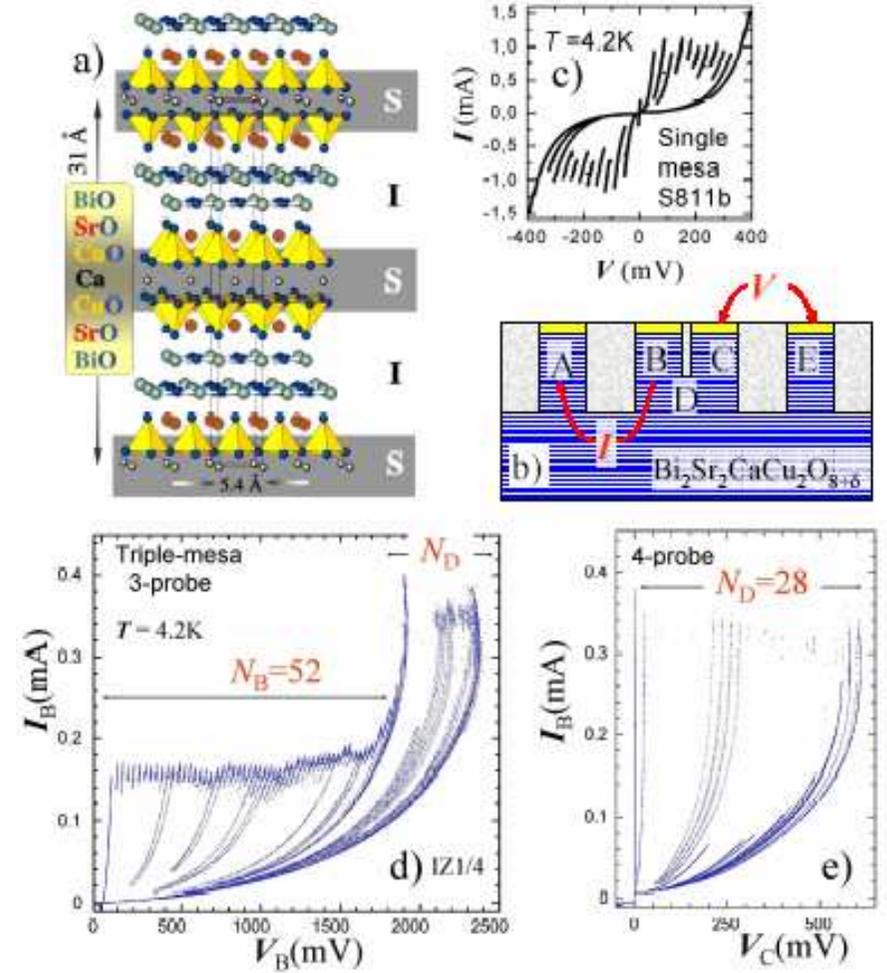

**Figure 1.** Crystallographic structure and *c*-axis *I-V* characteristics of Bi-2212.

a) The crystallographic structure of Bi-2212 (after A.Yurgens). Mobile charge carriers are confined to metallic ($CuO_2$-Ca-$CuO_2$) planes, while the *c*-axis transport is due to interlayer tunneling through blocking (SrO-2BiO-SrO) layers. Therefore, a single crystal is a stack of "intrinsic" tunnel junctions with the stacking periodicity 15.5 Å.

b) Sketch of a triple-mesa structure. Mesas B and C were obtained by making a narrow cut in the initial larger mesa D by Focused Ion Beam. Contact configuration used in four-probe measurements is shown.

c) IVC of a single overdoped mesa. Multi-branch structure is due to sequential switching of intrinsic Josephson junctions from the superconducting into the resistive state.

d) Three- and e) four-probe IVC's of a triple-mesa structure on optimally doped Bi-2212 single crystal at *T*=4.2K. Two families of QP branches, corresponding to junctions in mesas B and D, with ~0.45 difference in critical currents, are seen in the three-probe IVC. In the four-probe IVC only branches from the mesa D are seen.

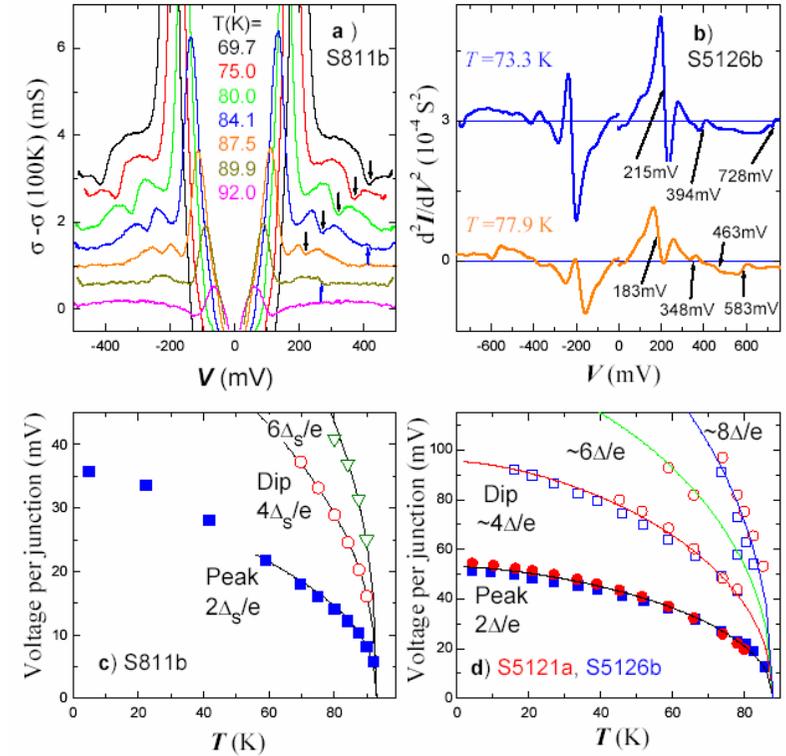

**Figure 2.** Temperature dependence of tunneling conductance of Bi-2212 mesas.

a) Tunneling conductance spectra for the same single mesa as in Fig 1 c). The peak at $V_g=2\Delta/e$ per IJJ is the dominant feature of the spectrum. Downwards and upwards arrows indicate main and secondary dips in $dI/dV$, respectively. Curves at different $T$ were shifted by 0.5 mS consecutively.

b) The second derivative $d^2I/dV^2(V)$ curves at two $T<T_c$ for an underdoped mesa. Curves at different $T$ were shifted for clarity.

c and d) $T$-dependences of the peak (solid symbols) and dips (open symbols) in $dI/dV$ are shown for c) the overdoped mesa and d) two mesas on the underdoped single crystal. The lines in both figures represent the fit to the peak voltage (the lower line) and multiple integers of this line. Coincidence of the lines with the open symbols indicate that dips occur at voltages, $V \approx 2(n+1)\Delta/e$ ($n=1, 2, 3$), per junction.

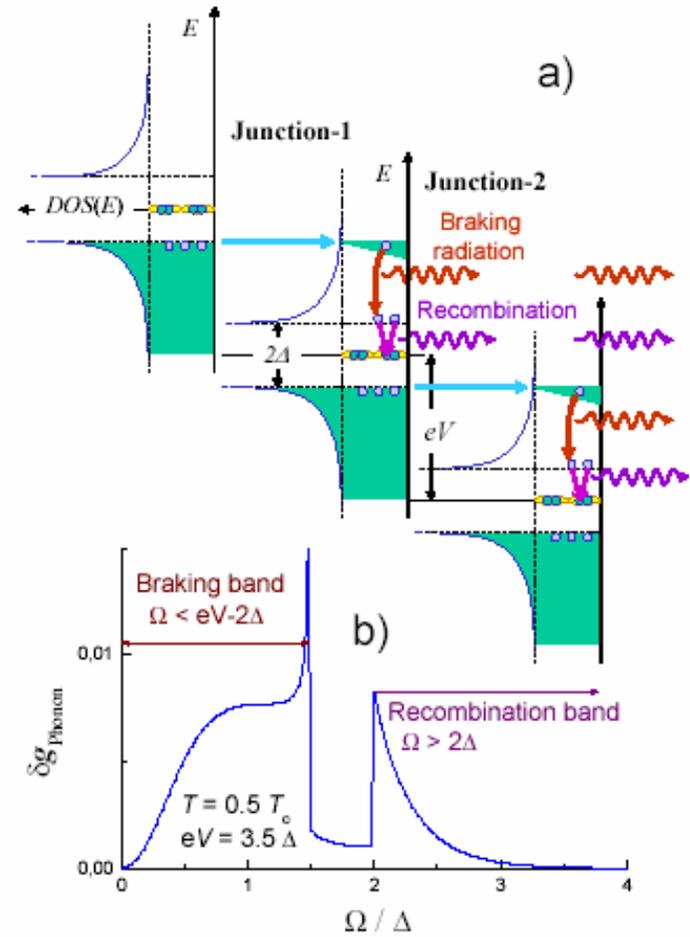

**Figure 3.** Mechanism of nonequilibrium quasiparticle relaxation in Josephson junctions.

a) Energy diagram (in a semiconductor representation) of two stacked SIS junctions biased at voltage $V$ per junction: For $V \geq 2\Delta/e$ tunneling results in nonequilibrium QP population in the empty band. Arrows indicate the most probable relaxation of nonequilibrium QP's. The process is repeated sequentially in the stacked junction-2, resulting in cascade amplification of radiation, as indicated by wavy arrows.

b) Calculated nonequilibrium part of phonon distribution function for a single SIS junction, biased at $V = 3.5\Delta/e$. Two phononic bands are clearly seen. The two bands overlap at $V = 4\Delta/e$, corresponding to the position of the first dip in $dI/dV(V)$ in Fig. 2.

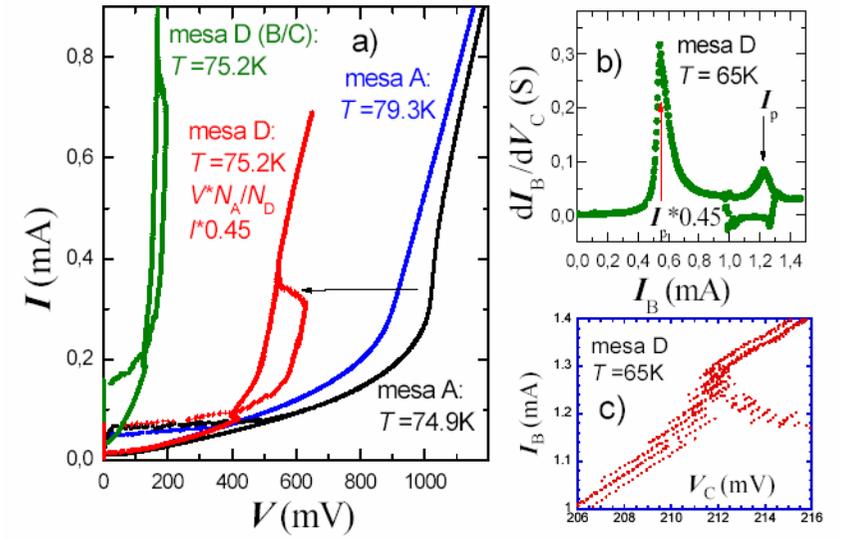

**Figure 4.** Generation-detection experiments with triple-mesa structures.

a) Four-probe IVC for the triple mesa D (green curve) and two IVC's for a single mesa A (black and blue curves) on the same crystal. A drop in the IVC of mesa D is seen at $I \sim 0.7$ mA. The red curve represents the same IVC of mesa D with voltage scaled by the number of junctions and current- by areas of mesas A and D. The arrow indicates that the drop occurs when IJJ's in mesa D reach the sum-gap bias.

b) Four-probe $dI_B/dV_C(I_B)$ characteristics for the triple mesa (subscripts indicate contact configurations, see Fig.1 b). Two peaks are observed, unlike for single mesa characteristics. The peak marked $I_p$ corresponds to the sum-gap bias in mesa D and is associated with the voltage drop ($dI/dV \approx 0$, at $I \sim 1.1$-$1.2$ mA). The larger peak at $I=0.45 I_p$ corresponds to the sum-gap bias of the mesa B. Those peculiarities measured at the unbiased detector mesa C indicate appearance of strong recombination radiation when IJJ's in either mesas B or D reach the sum-gap bias.

c) Enlarged part of the drop in the triple-mesa IVC at $T=65$K, demonstrating presence of tiny sub-branches, indicating that the non-equilibrium state in the stack is amplified in a cascade manner when IJJ's in mesa D sequentially reach the sum-gap bias.

## Appendix
**Calculation of non-equilibrium effects in superconducting tunnel junctions.**

### Definitions
In what follows we will use the following definitions:

$\delta N(E)$ and $\delta N(\Omega)$ are the number of nonequilibrium QP's and phonons at energy $E$ and $\Omega$, respectively.

$D_{QP}(0)$ - the quasiparticle density of states per spin at Fermi level.

$D_{Ph}(\Omega)$ - phonon density of states per unit sell (1/eV) $\approx a\Omega^2$ in the Debye approximation

$\alpha^2 D_{Ph}(\Omega)$ - electron-phonon spectral function $\approx b\Omega^2$

$N_I$ - density of ions (cm$^{-3}$).

$V$ - volume of electrodes, $V = Ad$, $A$ - area, $d$ – layer thickness.

$\Theta_D$ - Debye temperature, $\omega_D$ – Debye frequency.

$v_S$ - sound velocity.

$f(E) = \delta f(E) + F(E)$ and

$g(\Omega) = \delta g(\Omega) + G(\Omega)$ are nonequilibrium occupation numbers for QP's and phonons, respectively. Here $F(E)$ and $G(\Omega)$ are the corresponding equilibrium Fermi-Dirac and Bose-Einstein distribution functions.

$\rho(E) = \text{Re}\left[\dfrac{E - i\Gamma}{\sqrt{(E - i\Gamma)^2 - \Delta^2}}\right]$ is the normalized BCS QP density of states with a finite depairing factor $\Gamma$.

$\Delta$ is the superconducting energy gap,

$R_n$ - tunnel (normal) resistance of the junction;

$N_J$ – number of junctions in the stack.

### Kinetic balance equations
The non-equilibrium distribution of QP's and Phonons can be described by kinetic balance equations [17,20]:

$$\frac{\partial \delta N(E, \Omega)}{\partial t} = \frac{\partial \delta N}{\partial t}\bigg|_{(rel)} + \frac{\partial \delta N}{\partial t}\bigg|_{(inj)} + \frac{\partial \delta N}{\partial t}\bigg|_{(esc)}, \qquad (3)$$

which describes a dynamic equilibrium between relaxation (subscript "rel"), injection from nearby layer (subscript "inj"), and escape out of the junction (subscript "esc") of the particles.

The number of excess QP's in the energy interval from $E$ to $E+dE$ is
$$\delta N(E) = 2V\, D_{QP}(0)\rho(E)\delta f(E)dE.$$

The number of excess Phonons in the energy interval from $\Omega$ to $\Omega + d\Omega$ is
$$\delta N(\Omega) = V\, N_I\, D_{Ph}(\Omega)\delta g(\Omega)d\Omega$$

*QP relaxation rate*

$$\frac{\partial \delta N(E)}{\partial t}\bigg|_{(rel)} = -\frac{4\pi V\, D_{QP}(0)dE}{\hbar} \times$$

$$\int_0^\infty d\Omega\, \alpha^2 D_{Ph}(\Omega)\rho(E)\rho(E+\Omega)\left(1 - \frac{\Delta^2}{E(E+\Omega)}\right)\{f(E)[1-f(E+\Omega)]g(\Omega) - f(E+\Omega)[1-f(E)][1+g(\Omega)]\}$$

$$+ \int_0^{E-\Delta} d\Omega\, \alpha^2 D_{Ph}(\Omega)\rho(E)\rho(E-\Omega)\left(1 - \frac{\Delta^2}{E(E-\Omega)}\right)\{f(E)[1-f(E-\Omega)][1+g(\Omega)] - f(E-\Omega)[1-f(E)]g(\Omega)\}$$

$$+ \int_{E+\Delta}^\infty d\Omega\, \alpha^2 D_{Ph}(\Omega)\rho(E)\rho(\Omega-E)\left(1 + \frac{\Delta^2}{E(\Omega-E)}\right)\{f(E)f(\Omega-E)[1+g(\Omega)] - [1-f(E)][1-f(\Omega-E)]g(\Omega)\}$$

Here the first integral describes relaxation with absorption of a phonon, the second integral - relaxation with emission of a phonon and the third integral describes pair breaking and recombination.

*Phonon relaxation rate*

$$\frac{\partial \delta N(\Omega)}{\partial t}\bigg|_{(rel)} = -\frac{8\pi V\, D_{QP}(0)\alpha^2 D_{Ph}(\Omega)d\Omega}{\hbar} \times$$

$$\int_\Delta^\infty dE\, \rho(E)\rho(E+\Omega)\left(1 - \frac{\Delta^2}{E(E+\Omega)}\right)\{f(E)[1-f(E+\Omega)]g(\Omega) - f(E+\Omega)[1-f(E)][1+g(\Omega)]\}$$

$$+ \frac{1}{2}\int_\Delta^\infty dE\, \rho(E)\rho(\Omega-E)\left(1 + \frac{\Delta^2}{E(\Omega-E)}\right)\{[1-f(E)][1-f(\Omega-E)]g(\Omega) - f(E)f(\Omega-E)[1+g(\Omega)]\}$$

Here the first integral describes phonon emission/absorption due to QP relaxation/ excitation and the second integral – due to pairbreaking and recombination.

We can rewrite those two equations in terms of non-equilibrium occupation numbers $\delta f(E)$ and $\delta g(\Omega)$:

$$\frac{\partial \delta N(E)}{\partial t}\bigg|_{(rel)} = -\frac{4\pi V D_{QP}(0)\Delta^4 b}{\hbar}\frac{d(E)}{\Delta}\times$$

$$\int_0^\infty \frac{d\Omega}{\Delta}\frac{\Omega^2}{\Delta^2}\rho(E)\rho(E+\Omega)\left(1-\frac{\Delta^2}{E(E+\Omega)}\right)\{\delta f(E)[g(\Omega)+f(E+\Omega)]-\delta f(E+\Omega)[1-F(E)+g(\Omega)]+\delta g(\Omega)[F(E)-F(E+\Omega)]\}$$

$$+\int_0^{E-\Delta}\frac{d\Omega}{\Delta}\frac{\Omega^2}{\Delta^2}\rho(E)\rho(E-\Omega)\left(1-\frac{\Delta^2}{E(E-\Omega)}\right)\{\delta f(E)[1+g(\Omega)-f(E-\Omega)]-\delta f(E+\Omega)[F(E)+g(\Omega)]+\delta g(\Omega)[F(E)-F(E-\Omega)]\}$$

$$+\int_{E+\Delta}^\infty \frac{d\Omega}{\Delta}\frac{\Omega^2}{\Delta^2}\rho(E)\rho(\Omega-E)\left(1+\frac{\Delta^2}{E(\Omega-E)}\right)\{\delta f(E)[g(\Omega)+f(\Omega-E)]+\delta f(E+\Omega)[F(E)+g(\Omega)]+\delta g(\Omega)[F(E)+F(E+\Omega)-1]\}$$

(4)

$$\frac{\partial \delta N(\Omega)}{\partial t}\bigg|_{(rel)} = -\frac{8\pi V D_{QP}(0)\Delta^4 b}{\hbar}\frac{\Omega^2}{\Delta^2}\frac{d\Omega}{\Delta}\times$$

$$\int_\Delta^\infty \frac{dE}{\Delta}\rho(E)\rho(E+\Omega)\left(1-\frac{\Delta^2}{E(E+\Omega)}\right)\{\delta f(E)[g(\Omega)+f(E+\Omega)]-\delta f(E+\Omega)[1-F(E)+g(\Omega)]+\delta g(\Omega)[F(E)-F(E+\Omega)]\}$$

$$-\frac{1}{2}\int_\Delta^\infty dE\rho(E)\rho(\Omega-E)\left(1+\frac{\Delta^2}{E(\Omega-E)}\right)\{\delta f(E)[g(\Omega)+f(\Omega-E)]+\delta f(E+\Omega)[F(E)+g(\Omega)]+\delta g(\Omega)[F(E)+F(E+\Omega)-1]\}$$

(5)

Note that terms containing only equilibrium functions $F,G$ cancelled out, implying that the equilibrium state is not disturbed if there are no nonequilibrium QP's, $\delta f(E) = \delta g(\Omega) = 0$.

*QP injection rate*

QP's are injected via tunnelling from one electrode to another. Injection rate in the energy interval $E$ to $E+dE$ is

$$\frac{\partial \delta N(E)}{\partial t}\bigg|_{(inj)} = \frac{\Delta}{e^2 R_n}\frac{dE}{\Delta}\rho(E)\rho(E-eV)[f(E-eV)-f(E)].$$

(6)

*Phonon injection rate*

In a stack of junctions, nonequilibrium phonons which escape from one of the junctions enter the nearby junctions. Therefore, the injection rate can be estimated as

$$\frac{\partial \delta N(\Omega)}{\partial t}\bigg|_{(inj)} = -\beta N_J \frac{\partial \delta N(\Omega)}{\partial t}\bigg|_{(esc)}.$$

(7)

Here the factor $N_J$ indicates the cascade amplification of radiation in a stack and a factor $0<\beta<1$ describes the efficiency of such amplification.

*QP escape rate*

QP's can escape from an electrode via tunneling to another junction or into contact electrodes. The escape rate due to tunneling in the next junction can be written as

$$\frac{\partial \delta N(E)}{\partial t}\bigg|_{(esc)} = -\frac{\Delta}{e^2 R_n}\frac{dE}{\Delta}\rho(E)\rho(E+eV)[f(E)-f(E+eV)].$$

(8)

*Phonon escape rate*

Phonon escape rate is determined by the escape time $\tau \sim d/v_s$:

$$\frac{\partial \delta N(\Omega)}{\partial t}\bigg|_{(esc)} = -\delta N(\Omega)\frac{v_s}{d}$$

(9)

**Numerical procedure**

To solve the system of coupled integral Eqs. (3,4,6,8) and (3,5,7,9) we followed the procedure similar to Refs.[17,20]. Equations (4,5) for relaxation rates were linearized by substituting nonequilibrium $f(E)$ and $g(\Omega)$ under integrals by their equilibrium values $F(E)$ and $G(\Omega)$, which by definition can describe only slightly nonequilibrium states. Integrals in Eqs 4,5 were substituted by the corresponding sums for energy levels with separation $dE=\Delta/K_\Delta$ and the total number of levels $K$ for QP's and $K+2 K_\Delta$ for phonons:

$E_i = (i-0.5)dE, \quad (i=1,2,...K)$

$\Omega_i = (i-0.5)dE, \quad (i=1,2,...K+2K_\Delta)$

Thus the system of coupled integral equations Eq.3 was reduced to a system of linear equations [17]:

$R_{ij} x_j + \gamma_Y x_i = \gamma_I Y_i, \quad (1\leq i \leq K, \quad 1\leq j \leq 2K+2K_\Delta),$

$R_{ij} x_j + \gamma_U x_i = \gamma_P Y_i, \quad (K+1\leq i \leq K, \quad 1\leq j \leq 2K+2K_\Delta),$

(10)

where the first and second equations describe QP and phonon balance equations, respectively. Here

$x_i = \delta f(E_i), \quad (1\leq i \leq K),$

$x_i = \delta g(\Omega_i), \quad (K \leq i \leq 2K+2K_\Delta),$

The terms $R_{ij}x_j$ represent the relaxation equations (4,5), $\gamma_Y x_i$ and $\gamma_U x_i$ represent QP and phonon escape terms Eqs.(8,9) and right-hand side terms $\gamma_I Y_i$ and $\gamma_P Y_i$ represent the QP and phonon injection terms Eqs. (6,7), respectively. In the simulations presented here we used $K_\Delta=100$ and $K=900$, resulting in an energy interval $0<E<9\Delta$ for QP's and $0<E<11\Delta$ for phonons.

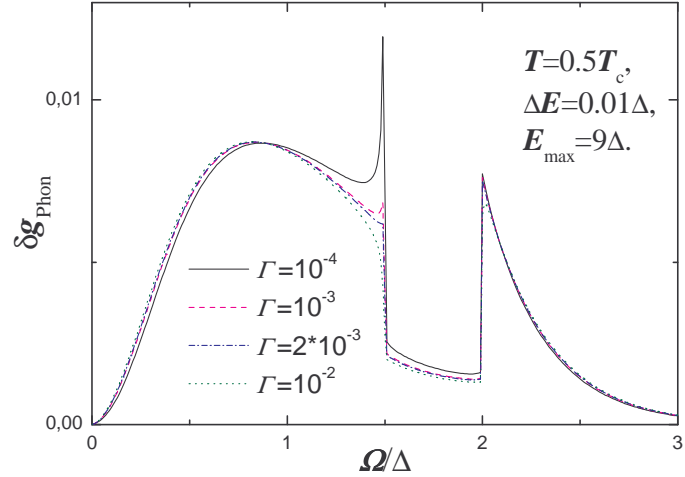

*Fig. S1. Simulated frequency dependence of non-equilibrium distribution functions of phonons at T=0.5T$_c$ for different depairing factors Γ.*

Fig. S1 shows simulated non-equilibrium distribution functions of phonons $T=0.5T_c$ for different depairing factors $\Gamma$, i.e. for different strength of singularities in the QP density of states at the edge of the gap. The presence of two phonon bands, the braking band $\Omega \approx eV-2\Delta$, and the recombination band $\Omega \approx 2\Delta$, are clearly seen. The states in between the bands are caused by finite temperature, $T=0.5T_c$, leading to a finite probability of excitation of tunneled QP's due to absorption of thermal (equilibrium) phonons.

As was noted in Ref.[16] the braking band exhibits a discontinuous edge at the maximum frequency which is mathematically due to the convolution of two BCS square root singularities in the QP density of states. From Fig. S1 it is seen that at low depairing factor (i.e. for strong square root singularity) the spectrum attains maxim at the edge of the braking band, which was not reported in previous simulations [16,20]. From Fig. S1 it is clear that this maximum is related to the strength of singularity at $E=\Delta$ and disappears at a certain depairing factor. In all simulations [16,17,20] (including our) certain smoothing of the singularity was required for convergence of the numerical procedure. It should be emphasized, that the simulations presented here are much more detailed than those presented before [16,17,20]. For example, the mesh size $\Delta E=0.01\Delta$ used in our simulations is much smaller than $\Delta/6$ and $\Delta/15$ used in Refs. [17] and [20]. Similarly, the frequency integration interval used here, $\Omega \approx 11\Delta$, is much larger than the corresponding intervals used in previous simulations. We observed that for smaller frequency intervals and larger mesh sizes the results of simulations were sensitive to mesh size and frequency interval. The absence of spurious effects in the numerical procedure was checked by changing the mesh size, $K_\Delta$, and the energy interval $K$. Mesh and interval sizes used in simulations presented here were chosen to avoid any spurious effects. Therefore, we don't see the reasons to believe that the maximum at the edge of the braking band for small $\Gamma$ (see the curve for $\Gamma = 10^{-4}$ in Fig. S1) is an artifact of the numerical procedure.

Anyway, it is the presence of the sharp edge (but not necessarily the maximum) which is important for appearance of stimulated emission at $V=4\Delta/e$. Therefore, in order to cope with the previous results [16], in Fig.3 b) we have chosen the curve for $\Gamma = 10^{-3}$, which has almost no maximum for the non-equilibrium distribution function $\delta g$ at the edge of the braking band. However, there is a clear maximum in the amount of non-equilibrium phonons $\delta N$, as seen in Fig 3 b), because it correspond to the product of $\delta g$ and the phononic density of states $D_{Ph}(\Omega)$. For obtaining Fig 3 b) we used Debye approximation $D_{Ph}(\Omega) = a\Omega^2$.